\input lanlmac
\input labeldefs.tmp
\writedefs
\input mssymb
\overfullrule=0pt

\font\email=cmtt9
%
\def\Z{{\bf Z}}
\def\MR#1{M_{R,#1}}
\def\MH#1{M_{H,#1}}
\def\MC#1{M_{C,#1}}
\def\MS#1{M_{S,#1}}
\def\Mc#1{M_{close,#1}}
\def\Mw#1{M_{wide,#1}}
\def\MCa#1{M_{C2,#1}}
\def\MCb#1{M_{C1,#1}}

\def\MCbd#1{M_{C1\downarrow,#1}}

\def\MCbu#1{M_{C1\uparrow,#1}}
\def\Mwd#1{M_{wide\downarrow,#1}}
\def\Mwu#1{M_{wide\uparrow,#1}}
\def\IH#1{I_{H,#1}}
\def\IC#1{I_{C,#1}}
\def\IS#1{I_{S,#1}}
\def\sgn{{\rm sign}(\pi-2\gamma)}
\def\th{\vartheta(\pi-2\gamma)}
\def\cox{h}
\def\Im{\mathop{\rm Im}\nolimits}
\def\Re{\mathop{\rm Re}\nolimits}
\def\sign{{\rm sign}}
\def\mod{\ {\rm mod}\ }
\def\neigh{{<\! st\! >}}
\def\der{\partial}
\def\to{\rightarrow}
\def\e#1{{\rm e}^{#1}}
%
\input epsf

\def\fig#1#2#3{
\xdef#1{\the\figno}
\writedef{#1\leftbracket \the\figno}
\nobreak
\par\begingroup\parindent=0pt\leftskip=1cm\rightskip=1cm\parindent=0pt
\baselineskip=11pt
\midinsert
#3
\vskip 12pt
{\bf Fig. \the\figno:} #2\par
\endinsert\endgroup\par
\goodbreak
\global\advance\figno by1
}
\Title{\vbox{\baselineskip12pt\hbox{LPTENS 97/65}\hbox{hep-th/9712222}}}
{\vbox{\centerline{Non-Linear Integral Equations}
\vskip2pt\centerline{for complex Affine Toda 
models}\vskip2pt\centerline{associated to simply laced Lie algebras}}}
 
\centerline{P.~Zinn-Justin\footnote{$^\dagger$}{{\email
pzinn@physique.ens.fr}}}
\medskip\centerline{Laboratoire de Physique Th\'eorique de l'Ecole
Normale Sup\'erieure\footnote{*}{unit\'e
propre du CNRS, associ\'ee \`a l'Ecole Normale
Sup\'erieure et l'Universit\'e Paris-Sud.}}
\centerline{24 rue Lhomond, 75231
Paris Cedex 05, France}
 
\vskip .3in
\centerline{Abstract}
A set of coupled non-linear integral (NLIE) equations is derived for a class of
models connected with the quantum group $U_q(\hat{\goth g})$ 
($\goth g$ simply laced Lie algebra), which are solvable using
the Bethe Ansatz; these equations describe arbitrary excited states
of a system with finite spatial length $L$. They generalize the
simpler NLIE of
the Sine--Gordon/massive Thirring model to affine Toda field
theory with imaginary coupling constant. As an application, the central
charge and all the conformal weights of the UV conformal field theory
are extracted in a straightforward manner. The quantum group
truncation for $q$ at a root of unity is discussed in detail;
in the UV limit we recover through this procedure the RCFTs
with extended $W(\goth g)$ conformal symmetry.

\Date{12/97 (revised 04/98)}

\newsec{Introduction}
The study of finite-size effects in 2D solvable lattice models (SLM) 
or integrable quantum field theories (IQFT) has proven to be a useful tool
to probe the physics of such systems. For critical SLM or conformal
field theories, one can extract from finite size-corrections the
central charge as well as all conformal weights
\ref\CAR{J.L.~Cardy, {\it Nucl.\ Phys.} B270 (1986), 186.}.
For IQFT which have non-trivial renormalization group flow (including
massive theories), one can obtain information on
the vicinity of the UV fixed point, of the IR fixed point, and the
crossover between the two. 

One possible way to investigate finite-size effects is the
Thermodynamic Bethe Ansatz equations \ref\YY{TBA equations were
originally introduced in\hfil\break
C.N.~Yang and C.P.~Yang,
{\it J.\ Math.\ Phys.} 10 (1969), 1115\semi
M.~Takahashi and M.~Suzuki, {\it Prog.\ Theor.\ Phys.} 48 (1972), 
2187\semi
and used in the context of finite-size effects in e.g.\hfil\break
Al.B.~Zamolodchikov, {\it Nucl.\ Phys.} B342 (1990), 695\semi
P.~Christe and M.~Martins, {\it Mod.\ Phys.\ Lett.} A5 (1990), 2189\semi
T.~Klassen and E.~Melzer, {\it Nucl.\ Phys.} B338 (1990), 485\semi
P.~Fendley and K.~Intriligator, {\it Nucl.\ Phys.} B372 (1992),
533.} which describe IQFT at finite
temperature $T$. This amounts to considering the euclidean
theory on a cylinder
with radius $\beta=1/T$. Modular invariance implies
that this can also be considered as the same theory at zero
temperature, but on a space which has been compactified with finite
length $L=\beta$. In this formulation,
the TBA only describe the ground state of the theory; in
particular, in the limit $T\rightarrow\infty$,
one can extract the central charge but not the conformal weights
of the UV conformal field theory, which correspond to low-lying
excited states.

Usually, in the TBA approach,
the central charge appears under the form of a dilogarithmic
sum, and it is known that one can extend these dilogarithmic
computations to obtain all conformal weights. So it is clear that
the TBA, after appropriate modification, may also yield excited
states \ref\KP{
TBA-like equations for excited states first appeared in\hfil\break
A.~Kl\"umper and P.~Pearce, {\it Physica} A183 (1992), 304\semi
for real TBA equations see for example\hfil\break
P.~Dorey and R.~Tateo, preprint hep-th/9706140.}.

Here another approach is used, which is related to the
methods of \ref\KLU{A.~Kl\"umper and M.~Batchelor, {\it J.\ Phys.}
A23 (1990), L189\semi
M.~Batchelor, A.~Kl\"umper and P.~Pearce,
{\it J.\ Phys.} A24 (1991), 3111\semi
A.~Kl\"umper, T.~Wehner and J.~Zittartz, {\it J.\ Phys.} A26 (1993),
2815\semi
More recent work includes:\hfil\break
G.~J\"uttner and A.~Kl\"umper, {\it Euro.\ Phys.\ Lett.} 37 (1997), 335\semi
G.~J\"uttner, A.~Kl\"umper and J.~Suzuki,
{\it Nucl.\ Phys.} B 487 (1997), 650\semi
G.~J\"uttner, A.~Kl\"umper and J.~Suzuki,
{\it J.\ Phys.} A30 (1997), 1181.}
and generalizes independant work by Destri and De~Vega
\nref\DDV{C.~Destri and H.J.~De~Vega, {\it Phys.\ Rev.\ Lett.} 69
(1992), 2313; {\it Nucl.\ Phys.} B438 (1995), 413.}
\nref\DDVexc{C.~Destri and H.J.~De~Vega,
{\it Nucl. Phys.} B504 (1997), 621.}[\xref\DDV--\xref\DDVexc].
The idea is to study directly
the Bethe Ansatz equations for an arbitrary state,
on a space of finite length $L$. The Bethe Ansatz are then
replaced with non-linear integral equations (NLIE)
which are much easier to handle; in particular they can be solved numerically. 
All relevant information on the system, including its energy, can then
be extracted from the NLIE. 

It is not clear at present how to write down NLIE for an
arbitrary integrable theory. The model investigated here is
the generalization of the massive Thirring/Sine--Gordon model,
which has $U_q(\widehat{\goth{sl}(2)})$ symmetry in the
$L\rightarrow\infty$ (infinite space) limit: we replace $\goth{sl}(2)$
with an arbitrary simply laced Lie algebra ${\goth g}=A_n$, $D_n$,
$E_{6,\, 7,\, 8}$ and consider the associated untwisted affine Lie algebra
$\hat{\goth g}$.
The continuous theory, which
is obtained as the scaling limit of an inhomogenenous
SLM -- the fact that we consider inhomogeneous transfer matrices
ensures the appearance of a mass gap in the theory -- is conjectured
to be Affine Toda Field theory $A_n^{(1)}$, $D_n^{(1)}$
or $E_{6,\, 7,\, 8}^{(1)}$ with imaginary coupling constant. We shall give
some strong arguments in favor of this hypothesis (mass spectrum and
scattering compatible with what has been conjectured before, correct UV limit).
We shall first write the standard
Bethe Ansatz equations using the Algebraic Bethe Ansatz, then
transform them into the NLIE, then finally take the scaling limit.
One should note that this scaling limit is not the same as the one
which leads to TBA equations. The TBA procedure involves two stages:
first sending the
spatial size $L$ to infinity in such a way that the density of B.A. roots and
the inverse lattice spacing 
remain of the order of the physical energy scale (which will eventually be
the temperature $T$);
then the inverse lattice spacing becomes the UV cutoff of the theory and is sent
to infinity keeping the mass scale $m$ of the order of $T$ ($m/T$ fixed).
On the contrary, we shall here keep $L$ finite: it will precisely define the
energy scale, and we shall introduce no other (there is no temperature $T$,
since we are considering the ground state and the low-lying excited states of
the theory). So there is only one stage,
which is to send the UV cutoff (the inverse lattice spacing) to infinity
while keeping $mL$ fixed.

As expected, for an algebra $\goth{g}$ of rank $n$, the NLIE form a set
of $n$ coupled equations labelled by a Dynkin diagram index.
However, the structure is still much simpler than
the corresponding TBA equations, which, due to the ``string
hypothesis'', are labelled by a second string index.
One consequence of the simple structure of the NLIE is that
the UV central charge and conformal weights will not appear as
infinite dilogarithmic sums (as in the TBA),
but as elementary finite sums.

One of the reasons
which make Affine Toda field theories with imaginary coupling
interesting
two-dimensional integrable field theories
is that, in spite of the non-hermiteanness of its Hamiltonian
\ref\COR{For a review see\hfil\break
E.~Corrigan, Lectures at the CRM-CAP Summer School 94, Banff,
Alberta, Canada (hep-th/9412213).} it shares many properties
of the Sine--Gordon theory: solitonic excitations (which are
expected to form $U_q(\goth{g})$ multiplets at the quantum level,
see \ref\BL{D.~Bernard and A.~Leclair, {\it Commun.\ Math.\ Phys.} 142 (1991),
99.}), breathers in the attractive regime \ref\HIM{U.~Harder,
A.~Iskandar and W.~McGhee, {\it Int.\ J.\ Mod.\ Phys.} A10 (1995), 1879.}.
It is also expected that one can consistently restrist these theories
at rational values of $\gamma/\pi$ ($q=-\e{-i\gamma}$) to yield 
(possibly unitary) theories which are perturbations of
$W(\goth g)$-symmetric rational conformal field theories
\ref\DVF{H.J.~De~Vega and V.A.~Fateev, {\it Int.\ J.\ 
Mod.\ Phys.} A6 (1991), 3221.}. All these points are discussed in
the present paper.

The article is organized as follows: In section 2, we review some basic
facts about the relevant lattice models and their Bethe Ansatz Equations.
In section 3, we study these equations and turn them into the NLIE.
Sections 4 and 5 are devoted to the computation of the energy/momentum
and of the $\goth{g}_0$ weight (which corresponds
to the $U_q(\goth g)$ representation in the $L\rightarrow\infty$ limit)
of a Bethe Ansatz state. In section 6 we briefly discuss the $L\rightarrow
\infty$ limit, whereas sections 7 and 8 are related to the $L\rightarrow 0$
limit and the computation of the UV central charge/conformal weights.
Section 8 explains how to perform the quantum group
truncation in the Bethe Ansatz formalism. This leads to considering
a restricted theory which is shown to be a perturbation of 
$W(\goth g)$-symmetric minimal models by studying its UV limit.
Finally, appendix A clarifies the intepretation of the UV spectrum,
explaining some technical issues which had so far remained
unclear even in the $A_1$ case.

\newsec{The lattice model and its Bethe Ansatz equations.}
Let $\goth g$ be a simply laced Lie algebra, and $\hat{\goth g}$ the
corresponding untwisted affine algebra.
We start with the $R$-matrix associated to
$U_q(\hat{\goth g})$
and the fundamental representation $V$ of $\goth g$.
For ${\goth g}=A_n$, we have: ($a,b=1\ldots n+1$, $a\ne b$)
\eqn\Rmat{\eqalign{
\check{R}^{aa}_{aa}(\Lambda)&=1\cr
\check{R}^{ab}_{ba}(\Lambda)&={\sin\Lambda\over\sin(\gamma-\Lambda)}\cr
\check{R}^{ab}_{ab}(\Lambda)&=
{\sin\gamma\over\sin(\gamma-\Lambda)}\e{i\Lambda\,\sign(a-b)}\cr
}}
$R$-matrices for other Lie algebras may be found in \ref\JIM{M.~Jimbo,
{\it Int.\ J.\ Mod.\ Phys.} A4, 15 (1989), 3759 and references therein.}.
$\gamma$ is the anisotropy parameter (which is related
to the deformation parameter $q$ of $U_q(\hat{\goth g})$ by $q=-\e{-i\gamma}$).
We define next the inhomogeneous transfer matrix $T(\Lambda,\Theta)$; it
is an operator in the physical Hilbert space ${\cal H}=V^{\otimes 2M}$ and it
depends on a spectral parameter $\Lambda$ and an inhomogeneity $\Theta$ (which
will eventually play the role of UV cutoff in spectral parameter space).
\eqn\Tmat{T(\Lambda,\Theta)=\tr_{aux}\left[
R_1(\Lambda-i\Theta) R_2(\Lambda+i\Theta) \ldots R_{2M-1}(\Lambda-i\Theta)
R_{2M}(\Lambda+i\Theta)\right]}
The $R$-matrix $R_i$ is simply the $R$-matrix acting on the tensor product
of the $i^{\rm th}$ component of $\cal H$ and of an auxiliary space $V_{aux}
\equiv V$ (with a permutation for correct labelling of the two spaces):
\eqn\Lmat{R_i(\Lambda)=\check{R}_{i,aux}(\Lambda){\cal P}_{i,aux}}
The trace in \Tmat\ is taken on the auxiliary space.
For ${\goth g}=A_1$, one can redefine the Boltzmann weights \Rmat\ to make
them real, and \Tmat\ reduces to the transfer matrix of the six-vertex
model, with anisotropy $\gamma$. Let us also mention that if one
removed the inhomogeneity $\Theta$, then one could also describe $T(\Lambda)$
as the generating function for commuting Hamiltonians in the $XXZ$ model and
its generalizations to higher rank algebras.

The diagonalization of $T(\Lambda,\Theta)$ leads to the so-called Algebraic
Bethe Ansatz \ref\ABA{The Algebraic Bethe Ansatz is done for $A_n$
in\hfil\break
O.~Babelon, H.J.~De~Vega and C.M.~Viallet, {\it Nucl.\ Phys.} B200 (1982), 266\semi
for $D_n$ in: H.J.~De~Vega and M.~Karowski, {\it Nucl.\ Phys.} B280 (1987), 225\semi
for $E_n$ it has not been done explicitly, one must use the
Analytic Bethe Ansatz, as in\hfil\break
N.Yu.~Reshetikhin, {\it Sov.\ Phys.\ JETP} 57 (1983), 691\semi
see also N.Yu.~Reshetikhin, {\it Lett.\ Math.\ Phys.} 14 (1987), 235.}.
First, one notes that $T(\Lambda,\Theta)$ commutes with the natural action
of the commutative Cartan algebra $\goth{g}_0$ on $\cal H$; therefore
all eigenstates can be chosen weight vectors, i.e.\ eigenvectors
of $\goth{g}_0$.

\fig\dynkin{Table of simply laced Lie algebras. Note that
$\dim{\goth g}=n(h+1)$.}{
$$\matrix{\hbox{\rm Lie algebra $\goth g$}& \hbox{\rm Dynkin diagram}&
\hbox{\rm Coxeter number $\cox$}& \dim{\goth g}\cr
\hrulefill&\hrulefill&\hrulefill&\hrulefill\cr
A_n&\vcenter{\vskip3mm\epsfbox{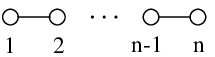}\vskip1mm} &n+1& n(n+2)\cr
D_n&\vcenter{\vskip3mm\epsfbox{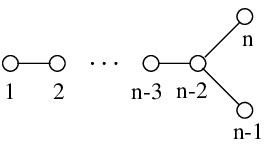}\vskip1mm}&2(n-1)&n(2n-1)\cr
E_6&\vcenter{\epsfbox{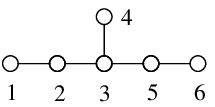}\vskip2mm}&12&78\cr
E_7&\vcenter{\epsfbox{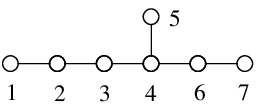}\vskip2mm}&18&133\cr
E_8&\vcenter{\epsfbox{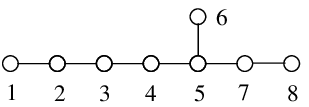}\vskip2mm}&30&248\cr
}$$
}

The eigenstates of $T$ are created from the
highest weight vector of $\cal H$ 
by the action of lowering operators which are
interpreted as creation operators of spin excitations. The latter are
labelled by a Dynkin diagram index $s=1\ldots n$ (see fig.~\dynkin)
and by a spectral parameter
(which we rescale by a factor $i\gamma h/2\pi$ and shall now call {\it rapidity})
$\lambda_{s,k}$ (where $k=1\ldots M_s$ runs over all spin excitations
of type $s$). As the model we consider is in an ``antiferromagnetic'' regime,
these excitations are not yet the physical excitations of the system. Their
rapidities satisfy the following set of coupled algebraic equations
(Nested Bethe Ansatz Equations):
\eqn\NBAE{\eqalign{
&\prod_{\scriptstyle j=1\atop\scriptstyle j\ne k}^{M_s}
{\sinh(\gamma(
{\cox\over 2\pi}(\lambda_{s,k}-\lambda_{s,j})+i))
\over\sinh(\gamma({\cox\over 2\pi}(\lambda_{s,k}-\lambda_{s,j})-i))}
\prod_{t|\neigh} \prod_{j=1}^{M_t}
{\sinh(\gamma(
{\cox\over 2\pi}(\lambda_{s,k}-\lambda_{t,j})-i/2))
\over\sinh(\gamma({\cox\over 2\pi}(\lambda_{s,k}-\lambda_{s,j})+i/2))}
\cr
&=\left[{\sinh(\gamma(
{\cox\over 2\pi}(\lambda_{s,k}-\theta)+i/2))
\over\sinh(\gamma({\cox\over 2\pi}(\lambda_{s,k}-\theta)-i/2))}
{\sinh(\gamma({\cox\over 2\pi}(\lambda_{s,k}+\theta)+i/2))
\over\sinh(\gamma({\cox\over 2\pi}(\lambda_{s,k}+\theta)-i/2))}
\right]^{M\delta_{s1}}
\cr}
}
where $\neigh$ means that $s$ and $t$ are neighbors on the Dynkin diagram of
$\goth{g}$. $\Theta$ has also been rescaled: $\Theta\equiv \gamma h\theta/2\pi$.
To each set of $\{ \lambda_{s,k} \}$ corresponds a Bethe Ansatz
state; calling $\e{-iE_\pm}$ the eigenvalue of $T$ for $\lambda=\pm\theta$, we have
\eqn\EP{\e{-iE_\pm}=\prod_{k=1}^{M_1} 
{\sinh(\gamma({\cox\over 2\pi}(\mp\lambda_{1,k}+\theta)+i/2))
\over\sinh(\gamma({\cox\over 2\pi}(\pm\lambda_{1,k}-\theta)+i/2))}
}
We are only interested in these particular values of $\lambda$ since in the scaling
limit (that will be defined later),
the energy $E$ and momentum $P$ can be extracted from them,
through the relation $E^\pm=(E\pm P)/2$, where $E^+$ and $E^-$
are given by \EP\ in inverse lattice spacing ($M/L$) units.
This relation can be derived in the ``light-cone approach'' \DDV.

Finally, the weight $r$ of the state with respect to $\goth g_0$
which can be decomposed as: 
$r=\sum_{s=1}^n r_s w_s$
on the basis of fundamental weights $w_s$
is given by
\eqn\rep{r_s=\delta_{s1}2M-\sum_{t=1}^n C_{st} M_t}
where $C_{st}$ is the Cartan matrix of $\goth{g}$: $C_{st}=2$ for $s=t$,
$-1$ for $\neigh$, $0$ otherwise.

For $\goth{g}=A_n$, $r_s$ is simply interpreted as the number of columns
of size $s$ in the Young tableau corresponding to $r$. 

\newsec{Derivation of the NLIE}
We shall now study solutions of the Bethe Ansatz Equations \NBAE\ 
which correspond to low-lying excited states (i.e.\ states with a finite
number of physical excitations above the vacuum);
we shall derive for each such solution a set of
Non-Linear Integral Equations, and then take the scaling limit.

\subsec{The counting functions.}
The basic quantities we need are the counting functions
$Z_s$, $s=1\ldots n$, defined by
\eqn\defZ{Z_s(\lambda) =
\delta_{s1} M (\phi_{1/2}(\lambda+\theta)+\phi_{1/2}(\lambda-\theta))
-\sum_{k=1}^{M_s} \phi_1(\lambda-\lambda_{s,k})
+\sum_{t|\neigh} \sum_{k=1}^{M_t} \phi_{1/2}(\lambda-\lambda_{t,k})}
where we have introduced the notation
\eqn\defphi{\phi_\alpha(\lambda)\equiv i\log
{\sinh(\gamma(+{\cox\over 2\pi}\lambda+i\alpha))
\over\sinh(\gamma(-{\cox\over 2\pi}\lambda+i\alpha))}}
The odd functions $\phi_\alpha$ are extended to the whole complex plane
by giving a prescription on their cuts (see fig. \cuts; this is the
same convention as in \DDVexc)
\fig\cuts{Cuts of the function $\phi_\alpha(\lambda)$
for $\alpha<{\pi\over 2\gamma}$
(left) and $\alpha>{\pi\over 2\gamma}$ (right). Arrows correspond to
jumps of $+2\pi$ of the function.}{\centerline{\epsfbox{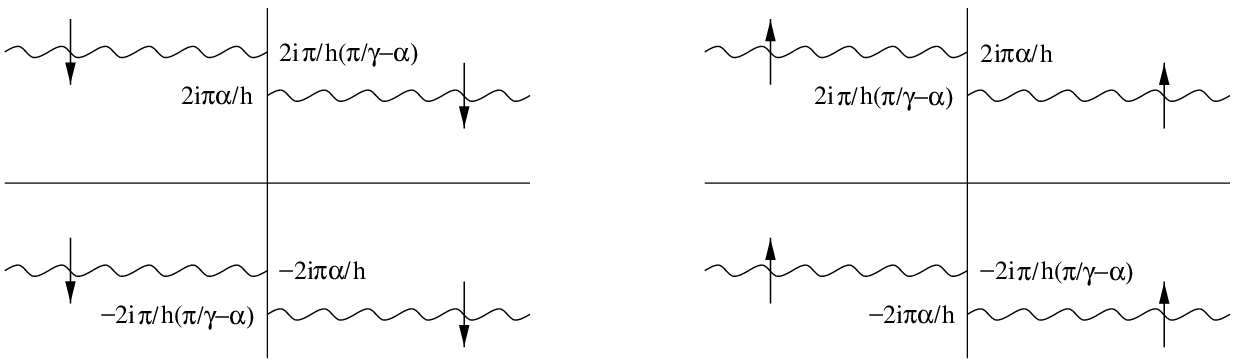}}}

The key property of $Z_s$ is that, according to BAE \NBAE,
for each root $\lambda_{s,k}$ of the BAE, we have:
\eqn\prop{Z_s(\lambda_{s,k})=2\pi I_{s,k}}
where one can check that $I_{s,k}$ is a half-integer, whose parity (i.e.\ 
$2I_{s,k}\mod 2$) is the opposite of that of $r_s+M_s$.

It should be noted that property \prop\ is true not only for real roots
but also for complex roots (since $Z_s$ has been defined on the whole
complex plane), in contrast to what is usually done when writing
Bethe Ansatz Equations in the thermodynamic limit.

Now let us classify the diffent types of roots that appear; we shall
restrict ourselves to configurations of roots which survive in the
scaling limit.

$\bullet$ {\sl Real roots and holes.}

Since it is known that the ground state consists
of real roots of all types ($s=1\ldots n$), we expect that for an arbitrary
low-lying excited state, we shall have a large number of real roots
(divergent in the thermodynamic limit) that we denote by $\rho_{s,k}$
($k=1\ldots \MR{s}$).
If one considered the ground state, the real roots would in fact exhaust
all the half-integer values (with appropriate parity) 
of $Z_s(\lambda)/2\pi$ with $\lambda$ real; for an
excited state, on the other hand, there may be a finite number of
real $\lambda$ which are
distinct from all $\rho_{s,k}$ but still satisfy this property. We call these
holes and write
\eqn\propb{Z_s(\eta_{s,k})=2\pi \IH{s,k}\qquad k=1\ldots \MH{s}}

$\bullet$ {\sl Special roots/holes.}

Because of the driving term
$\delta_{s1} M (\phi_{1/2}(\lambda+\theta)+\phi_{1/2}(\lambda-\theta))$,
which acts on equation $s=1$ and is transmitted to all equations
by the nearest-neighbor interaction on the Dynkin diagram, it is clear
that $Z_s$ must in general be an increasing function on the real axis. In fact
for a thermodynamic state (for example at finite temperature, when the number
of excitations is large) this statement is certainly true. However, it was
pointed out in \DDVexc\ that for low-lying excited states, there
might be local variations of $Z_s$ due to isolated roots, so that $Z_s$ 
is decreasing on a small interval. This behavior may become important if
$Z_s$ decreases enough to cross again $2\pi$ times a half-integer. We 
therefore introduce real parameters
$\sigma_{s,k}$ satisfying
\eqn\propc{Z_s(\sigma_{s,k})=2\pi \IS{s,k}\qquad k=1\ldots \MS{s}}
and $Z'_s(\sigma_{s,k})<0$.
The $\sigma_{s,k}$ may or may not be roots of the B.A.E.: they are called
respectively special roots and special holes.

$\bullet$ {\sl Complex roots.}

Because of the $2\pi^2/\cox\gamma$-periodicity of the equations, one can
assume that $|\Im \lambda_{s,k}|\le \pi^2/\cox\gamma$. With this convention,
all non-real roots of the B.A.E. will be called complex roots and
denoted by $\xi_{s,k}$, $k=1\ldots \MC{s}$. 
Further distinctions must be introduced to classify the complex roots.
We shall treat together the two regimes: repulsive regime for $\gamma<\pi/2$
and attractive regime for $\gamma>\pi/2$.

The first classification is the following:
$\xi_{s,k}$ is called a wide root if
$|\Im\xi_{s,k}|>\min(2\pi/\cox,2\pi/\cox(\pi/\gamma-1))$, a close root otherwise.
There are $\Mw{s}$ wide roots of type $s$ and $\Mc{s}$ close roots.

We also define a second, independent classification: 
$\xi_{s,k}$ is called of the first kind if $|\Im\xi_{s,k}|>\pi/\cox$,
of the second kind otherwise. We call $\MCb{s}$ (resp. $\MCa{s}$) the
numbers of roots of the first (resp. second) kind.

To clarify these definitions, we notice that there are three cases depending
on the value of $\gamma$.
In the repulsive regime ($\gamma<\pi/2$), all wide roots
are of the first kind, whereas a close root $\xi$ can be either
of the first kind ($\pi/\cox<|\Im\xi|<2\pi/\cox$) or of the second kind
($0<|\Im\xi|<\pi/\cox$). In the ``weakly attractive'' regime 
($\pi/2<\gamma<2\pi/3$), again all wide roots are of the first kind,
and there still are close roots of the first kind ($\pi/\cox<|\Im\xi|<
2\pi/\cox(\pi/\gamma-1)$). Finally, in the ``strongly attractive'' regime,
all close roots are of the second kind, and wide roots are either
of the second kind ($2\pi/\cox(\pi/\gamma-1)<|\Im\xi|<\pi/\cox$) or of the
first kind ($|\Im\xi|>\pi/\cox$).

\subsec{The NLIE.}
The derivation is a straightforward generalization of \DDVexc. 
We first assume that there are no special roots/holes.
We then use the following trick: as the
real zeroes of the function $1+(-1)^{\delta_s}\e{iZ_s(z)}$ 
($\delta_s\equiv r_s+M_s \mod 2$)
are exactly the real roots and the holes of type $s$, we have
\eqn\trick{\oint_C {dz\over 2\pi i} f(z) {d\over dz} 
\log(1+(-1)^{\delta_s}\e{iZ_s(z)}) = \sum_{k=1}^{\MR{s}} f(\rho_{s,k})
+\sum_{k=1}^{\MH{s}} f(\eta_{s,k})}
for an arbitrary analytic function $f$. The contour $C$ is a closed curve
which encircles all the $\rho_{s,k}$ and $\eta_{s,k}$. 

We apply \trick\ to the definition \defZ\ of $Z_s(\lambda)$ 
($\lambda$ real) differentiated once: we obtain
\eqn\dera{\eqalign{
Z'_s(\lambda) =
\delta_{s1} 2\pi M &(\Phi_{1/2}(\lambda+\theta)+\Phi_{1/2}(\lambda-\theta))\cr
-\bigg[&
\oint_C {dz\over i} \Phi_1(\lambda-z) 
{d\over dz} \log (1+(-1)^{\delta_s} \e{iZ_s(z)})\cr
&-2\pi \sum_{k=1}^{\MH{s}} \Phi_1(\lambda-\eta_{s,k})
+2\pi \sum_{k=1}^{\MC{s}} \Phi_1(\lambda-\xi_{s,k})
\bigg]\cr
+\sum_{t|\neigh} \bigg[&
\oint_C {dz\over i} \Phi_{1/2}(\lambda-z)
{d\over dz} \log (1+(-1)^{\delta_t} \e{iZ_t(z)})\cr
&-2\pi \sum_{k=1}^{\MH{t}} \Phi_{1/2}(\lambda-\eta_{t,k})
+2\pi \sum_{k=1}^{\MC{t}} \Phi_{1/2}(\lambda-\xi_{t,k})
\bigg]\cr
}}
where we have introduced $\Phi_\alpha=
(1/2\pi)d\phi_\alpha/d\lambda$. Next we deform the contour $C$ so that
the contour integrals can be rewritten as integrals on the real axis:
\eqn\derb{\eqalign{
Z'_s(\lambda) =
\delta_{s1} 2\pi M &(\Phi_{1/2}(\lambda+\theta)+\Phi_{1/2}(\lambda-\theta))\cr
-\bigg[&
\int dx\, \Phi_1(\lambda-x) Z'_s(x)
+\int dx\, \Phi_1(\lambda-x) {1\over i}{d\over dx} \log 
{(-1)^{\delta_s}+\e{-iZ_s(x-i0)}\over 1+(-1)^{\delta_s} \e{iZ_s(x+i0)}}\cr
&-2\pi \sum_{k=1}^{\MH{s}} \Phi_1(\lambda-\eta_{s,k})
+2\pi \sum_{k=1}^{\MC{s}} \Phi_1(\lambda-\xi_{s,k})
\bigg]\cr
+\sum_{t|\neigh} \bigg[&
\int dx\, \Phi_{1/2}(\lambda-x) Z'_t(x)
+\int dx\, \Phi_{1/2}(\lambda-x) {1\over i}{d\over dx} \log 
{(-1)^{\delta_t}+\e{-iZ_t(x-i0)}\over 1+(-1)^{\delta_t} \e{iZ_t(x+i0)}}\cr
&-2\pi \sum_{k=1}^{\MH{t}} \Phi_{1/2}(\lambda-\eta_{t,k})
+2\pi \sum_{k=1}^{\MC{t}} \Phi_{1/2}(\lambda-\xi_{t,k})
\bigg]\cr
}}
This equation suggests the introduction of the real function
\eqn\defQ{Q_s(x)={1\over i} \log
{1+(-1)^{\delta_s} \e{iZ_s(x+i0)}
\over 
1+(-1)^{\delta_s}\e{-iZ_s(x-i0)}}
}
(note that $Q_s$ is {\it a priori} defined up to a constant since only its
derivative appears in \derb, but the definition above turns
out to be convenient). $Q_s$ clearly satisfies 
\eqn\defQb{Q_s(x)=(Z_s(x)+\delta_s\pi) \mod 2\pi}
so that it is entirely
defined on the real axis (by appropriate choice of the logarithmic
cuts) by imposing \defQb\ and $|Q_s(x)|\le \pi$.

To simplify \derb\ we introduce the spectral-parameter
dependent Cartan matrix $C_{st}(\lambda)$:
\eqn\defC{C_{st}(\lambda)\equiv\left\{\eqalign{
2\delta(\lambda)\quad &s=t\cr
-{\cox\over 2\pi} {1\over\cosh(\cox\lambda/2)}
\equiv-2s(\lambda)
\quad&\neigh\cr
}\right.}
for $1\le s,t\le n$. In the sequel we shall also use Fourier transform
defined by
\eqn\defour{f(\kappa)=\int d\lambda\, \exp(i\kappa\,\cox\lambda/\pi) 
f(\lambda)}
for any function $f$.
With this convention $s(\kappa)=1/(2\cosh(\kappa))$. Note in particular
that $C_{st}(\kappa=0)\equiv C_{st}$
is the usual Cartan matrix of $\goth g$.

We now rewrite \derb:
\eqn\derc{\eqalign{
\sum_{t=1}^n C_{st} \star (1+\Phi_1)\star Z'_t(\lambda)&=
\delta_{s1} 4\pi M (\Phi_{1/2}(\lambda+\theta)+\Phi_{1/2}(\lambda-\theta))\cr
&+\sum_{t=1}^n \bigg[
(C_{st}\star(1+\Phi_1)-2\delta_{st})\star{d\over d\lambda}Q_t\cr
&-2\pi \sum_{k=1}^{\MH{t}} (C_{st}\star(1+\Phi_1)-2\delta_{st}) (\lambda-\eta_{t,k})\cr
&+2\pi \sum_{k=1}^{\MC{t}} (C_{st}\star(1+\Phi_1)-2\delta_{st}) (\lambda-\xi_{t,k})
\bigg]\cr
}}
where $\star$ means convolution product in $\lambda$ space, and $1$ is
the identity operator (convolution with the $\delta$ function).

We multiply by the inverse matrix $C^{-1}_{st}\star (1+\Phi_1)^{-1}$, and
then take the scaling limit $M\rightarrow\infty$,
$\theta\rightarrow\infty$ keeping $mL\equiv M \e{-\theta}$ fixed.
Expanding the inhomogeneous term $4\pi C^{-1}_{s1}\star(s(\lambda+\theta)
+s(\lambda-\theta))$ as $\theta\rightarrow\infty$, one finds
\ref\ORW{E.~Ogievietski and P.B.~Wiegmann, {\it Phys.\ Lett.} B168
(1986), 360\semi
N.Y.~Reshetikhin and P.B.~Wiegmann, {\it Phys.\ Lett.} B189 (1987),
125.} that the Perron--Frobenius eigenvalue of the Cartan matrix
(or, more precisely, of the adjacency matrix of its Dynkin
diagram) dominates, so that
\eqn\derd{Z'_s=m_s L \cosh \lambda
+\sum_{t=1}^n \left[ X_{st} \star {d\over d\lambda} Q_t
+\sum_{k=1}^{\MH{t}} X_{st}(\lambda-\eta_{t,k})
-\sum_{k=1}^{\MC{t}} X_{st}(\lambda-\xi_{t,k}) \right]}
The $m_s$ are the masses
of the solitons of the theory; they form the Perron--Frobenius
eigenvector, and they are of the order of the
mass scale $m$. 
The $X_{st}$ are regular functions; on the real axis they are given
by
$X_{st}(\lambda)=\delta_{st}\delta(\lambda)-2(1+\Phi_1)^{-1}\star
C_{st}^{-1}$, so that
their Fourier transforms are:
\eqn\XFour{X_{st}(\kappa)=\delta_{st}-{\sinh({\pi\over\gamma} \kappa)
\over \sinh(({\pi\over\gamma}-1) \kappa) \cosh(\kappa)} C^{-1}_{st}(\kappa)}
where $C^{-1}_{st}(\kappa)$ is listed in fig. \table\ for the infinite
series $A_n$, $D_n$.
\fig\table{Table of inverse Cartan matrices $C_{st}^{-1}(\kappa)$
and of the Perron--Frobenius eigenvectors (which give the mass spectrum).}{
$$\matrix{\hbox{$\goth g$}& C_{st}^{-1}(\kappa),\,s\ge t&
\hbox{\rm P-F. eigenvector}\cr
\hrulefill&\hrulefill&\hrulefill\cr
&&\cr
A_n& \coth(\kappa) {\displaystyle\sinh((n+1-s)\kappa)\sinh(t\kappa)
\over\displaystyle\sinh((n+1)\kappa)}& \sin\left({\displaystyle \pi s\over\displaystyle n+1}\right)\cr
&&\cr
D_n& \left\{\eqalign{
      &\coth(\kappa) {\displaystyle\cosh((n-1-s)\kappa)\sinh(t\kappa)
      \over\displaystyle\cosh((n-1)\kappa))}\quad s\le n-2\cr
      &\coth(\kappa) {\displaystyle\sinh(t\kappa)
      \over\displaystyle 2\cosh((n-1)\kappa))}\ s\ge n-1,\ t\le n-2 \cr
      &{\displaystyle\sinh(n\kappa)\over 2\sinh(\kappa)\cosh((n-1)\kappa)}
      \quad s=t\ge n-1\cr
      &{\displaystyle\sinh((n-2)\kappa)\over 2\sinh(\kappa)\cosh(n-1)\kappa)}
      \quad s=n,\ t=n-1}\right. 
&\left\{\eqalign{
      &\sin\left({\pi s\over 2(n-1)}\right)\ s\le n-2\cr
      &{1\over 2}\quad s=n-1,\, n}\right.
}$$}

However, to define $X_{st}(\lambda-\xi_{t,k})$ \derd\ one must extend
$X_{st}$ to the complex plane; one must then be careful that the poles of
the functions $\Phi_1$ and $\Phi_{1/2}$ are smeared by the convolution
product with $C^{-1}_{st}\star (1+\Phi_1)^{-1}$ and become cuts
running parallel to the real axis. In other words, $X_{st}$ is not
simply the analytic continuation $X_{st}^{(0)}$ of its
definition on the real axis; rather, it is given by:
\eqn\defX{\eqalign{
X_{st}(\lambda)=X_{st}^{(0)}(\lambda)
&+\vartheta\left(\Im\lambda-{2\pi\over h}\right)
X_{st}^{(0)}\left(\lambda-i{2\pi\over h}\right)\cr
&-\vartheta\left(\Im\lambda-{2\pi\over h}(\pi/\gamma-1)\right)
X_{st}^{(0)}\left(\lambda-i{2\pi\over h}(\pi/\gamma-1)\right)\cr
&-\vartheta\left(\Im\lambda-{\pi\over h}\right) 
\sum_{t'|<\! tt'\! >} X_{st'}^{(0)}\left(\lambda-i{\pi\over h}\right)\cr
}}
for $\Im\lambda>0$ (and a similar expression for
$\Im\lambda<0$). $\vartheta$ is the usual step function.

Finally one can safely integrate once \derd\ 
(taking care of the integration constant, which vanishes
since it has been absorbed in the definition \defQ\ of $Q_s$),
and reintroduce the special
roots/holes as in \DDVexc. The final equation is:
\eqn\dere{\eqalign{
Z_s= m_s L \sinh \lambda
+\sum_{t=1}^n \Bigg[ &X_{st} \star Q_t
+\sum_{k=1}^{\MH{t}} \chi_{st}(\lambda-\eta_{t,k})\cr
&-2\sum_{k=1}^{\MS{t}} \chi_{st}(\lambda-\sigma_{t,k})
-\sum_{k=1}^{\MC{t}} \chi_{st}(\lambda-\xi_{t,k}) \Bigg]}
}
with $\chi_{st}$ the odd primitive of $2\pi X_{st}$ (for
$\chi_{st}(\lambda
-\xi_{t,k})$ one should integrate on a line parallel to the real axis).

\newsec{Energy and momentum in the scaling limit.}
To each set of counting functions $Z_s$ that satisfy
the NLIE \dere\ is associated a 
configuration of holes and complex roots
which characterizes the corresponding excited state.
Conversely, specifying the approximate positions of holes and
complex roots\foot{The positions of holes cannot not be chosen arbitrarily
since they are ``quantized'' at finite $L$. The quantization
condition itself depends on positions of other holes and complex
roots because of the interaction between the roots.},
one can solve the 
non-linear equations \dere\ (at least numerically) and obtain the
counting functions $Z_s(\lambda)$. We shall now go on
and show how to express the energy/momentum in terms of the $Z_s$.
We shall not give all the details of the derivation since it is very
similar to the derivation of the NLIE itself. We introduce the
auxiliary function
\eqn\W{W(\lambda)=\sum_{k=1}^{M_1} \Phi_{1/2}(\lambda-\lambda_{1,k})}
and use the contour integral trick to express it as
\eqn\Wb{\eqalign{
W(\lambda)&=\oint_C {dz\over 2\pi i} \Phi_{1/2}(\lambda-z)
{d\over dz} \log(1+(-1)^{\delta_1}\e{iZ_1(z)})\cr
&-\sum_{k=1}^{\MH{1}}\Phi_{1/2}(\lambda-\eta_{1,k})
+2\sum_{k=1}^{\MS{1}}\Phi_{1/2}(\lambda-\sigma_{1,k})
+\sum_{k=1}^{\MC{1}} \Phi_{1/2}(\lambda-\xi_{1,k})
}}
then use the NLIE:
\eqn\Wc{W(\lambda)=W_0(\lambda)-\sum_{s=1}^n 
\left[{1\over 2\pi} G_s\star Q'_s
 -\sum_{k=1}^{\MH{s}} G_s(\lambda-\eta_{s,k})
+2\sum_{k=1}^{\MS{s}} G_s(\lambda-\sigma_{s,k})
+\sum_{k=1}^{\MC{s}} G_s(\lambda-\xi_{s,k})\right]
}
$W_0(\lambda)$ is a function that plays no role and will contribute to
the ground state bulk energy. $G_s(\lambda)\equiv 2C^{-1}_{s1}\star
s(\lambda)$ on the real axis, but like the $X_{st}$ its definition
differs in the complex plane. We shall not bother to write down the analogue
of Eq. \defX\ for $G_s$, since we are only interested in the scaling limit
($\theta$ and $M\rightarrow\infty$), in which this discussion
simplifies drastically. Indeed, the expansion of
$G_s(\lambda\pm\theta)$ (and the use of the Perron--Frobenius eigenvector
property $\sum_{t|\neigh} m_t = 2\cos(\pi/\cox) m_s$)
leads to $2\pi{M\over L}G_s(\lambda-\theta)
\sim {1\over 2} m_s e(\lambda)$ where 
\eqn\defe{\eqalign{
e(\lambda)\equiv \e\lambda \bigg(1&+
\vartheta\left(\Im\lambda-{2\pi\over \cox}\right)\e{-2i\pi/\cox}\cr
&-\vartheta\left(\Im\lambda-{2\pi\over\cox}(\pi/\gamma-1)\right)
\e{-2i\pi/\cox(\pi/\gamma-1)}\cr
&-\vartheta\left(\Im\lambda-{\pi\over \cox}\right)
(1+\e{-2i\pi/\cox})
\bigg)}
}
(for $\Im\lambda>0$). 

After integrating once $W(\lambda)$
and plugging \Wc\ in the definition \EP\ of the energy, we find:
\eqn\Escal{\eqalign{
E=\sum_s m_s \bigg[ &\sum_{k=1}^{\MH{s}} \cosh \eta_{s,k}
-2\sum_{k=1}^{\MS{s}} \cosh \sigma_{s,k}
-\sum_{k=1}^{\MC{s}} {1\over 2}(e(\xi_{s,k})+e(-\xi_{s,k}))\cr
&-{1\over 2\pi} \int d\lambda \, \cosh \lambda \, Q_s(\lambda) \bigg]
}}
where we have discarded the bulk ground state energy.
We shall now discuss in more detail the contribution of the complex roots,
which we call $E_C$ and redecompose: $E_C=E^+_C+E^-_C$. We
treat separately the repulsive and attractive regimes.

When $\gamma<\pi/2$ (repulsive regime), according to \defe,
the contribution of the wide roots to
$W(\lambda)$ and $E$ vanishes. The close roots do contribute:
\eqn\ECa{E^+_C={1\over 2}\sum_{s=1}^n m_s \Bigg[
-\hskip-10pt\sum_{k=1\atop\ |\Im\xi_{s,k}|<\pi/\cox} \hskip-10pt
\e{\xi_{s,k}}\ +
\hskip-15pt\sum_{k=1\atop\ \pi/\cox<|\Im\xi_{s,k}|<2\pi/\cox}\hskip-15pt
\e{\xi_{s,k}-2i\pi/\cox\, \epsilon_{s,k}}
\Bigg]
}
where $\epsilon_{s,k}\equiv \sign(\Im\xi_{s,k})$.

When $\gamma>\pi/2$ (attractive regime), all complex roots contribute;
if $\pi/2<\gamma<2\pi/3$ we find
\eqn\ECb{\eqalign{
E^+_C={1\over 2}\sum_{s=1}^n m_s \Bigg[
&-\hskip-10pt\sum_{k=1\atop\ |\Im\xi_{s,k}|<\pi/\cox}^n \hskip-10pt
\e{\xi_{s,k}}\ +
\hskip-20pt\sum_{k=1\atop\ \pi/\cox<|\Im\xi_{s,k}|<2\pi/\cox(\pi/\gamma-1)}^n
\hskip-20pt
\e{\xi_{s,k}-2i\pi/\cox\, \epsilon_{s,k}}\cr
&+
\hskip-15pt\sum_{k=1\atop|\Im\xi_{s,k}|>2\pi/\cox(\pi/\gamma-1)}^n\hskip-15pt
\e{\xi_{s,k}}(\e{-2i\pi/\cox\, \epsilon_{s,k}}+\e{-2i\pi/\cox(\pi/\gamma-1)
\, \epsilon_{s,k}})
\Bigg]
}}
whereas for $\gamma>2\pi/3$ we find
\eqn\ECc{\eqalign{
E^+_C={1\over 2}\sum_{s=1}^n m_s \Bigg[
&-\hskip-15pt\sum_{k=1\atop\ |\Im\xi_{s,k}|<2\pi/\cox(\pi/\gamma-1)}^n \hskip-15pt
\e{\xi_{s,k}}\ +
\hskip-20pt\sum_{k=1\atop\ 2\pi/\cox(\pi/\gamma-1)<|\Im\xi_{s,k}|<\pi/\cox}^n
\hskip-20pt
\e{\xi_{s,k}}(-1+\e{-2i\pi/\cox(\pi/\gamma-1)\, \epsilon_{s,k}})\cr
&+
\hskip-10pt\sum_{k=1\atop|\Im\xi_{s,k}|>\pi/\cox}^n\hskip-10pt
\e{\xi_{s,k}}(\e{-2i\pi/\cox\, \epsilon_{s,k}}+\e{-2i\pi/\cox(\pi/\gamma-1)
\, \epsilon_{s,k}})
\Bigg]
}}

\newsec{Relation between numbers of holes/complex roots and representation
of the state.}
It is convenient to express now the weight $r$ of the low-lying
excited states in terms of quantities which remain finite when we take the
scaling limit; indeed, \rep\ expresses $r$ in terms of the $M_s$ which
diverge as $M\rightarrow\infty$. Instead we shall derive now
a relation between $r$ and the numbers of holes, special roots/holes and
complex roots.

We start by considering the limit $\lambda\rightarrow+\infty$ in
the definition \defZ\ of $Z_s$: we obtain
\eqn\Zinf{\eqalign{
Z_s(+\infty)&=\delta_{s1} (\pi-\gamma) 2M - (\pi - 2\gamma) M_s
+\sum_{t|\neigh} (\pi-\gamma) M_t\cr
&+ 2\pi \,\sgn \Mwd{s} -2\pi\sum_{t|\neigh} \MCbd{t}\cr
&= (\pi-\gamma) r_s + \pi M_s \cr
&+ 2\pi \,\sgn \Mwd{s} -2\pi\sum_{t|\neigh} \MCbd{t}\cr
}}
and a similar expression for $Z_s(-\infty)$. The sign $\downarrow$ indicates
that we are counting the number of roots $\xi$ which satisfy $\Im\xi<0$.
In principle, we have $\Mwd{s}={1\over 2} \Mw{s}$ and $\MCbd{s}={1\over 2}
\MCb{s}$ since complex roots come in conjugate pairs; however we do not
need these relations.

Next we count the number of integer values of $Z_s$ on the real axis.
For this purpose we introduce
$I^{max}_s$ (resp. $I^{min}_s$), which
is the largest (resp. smallest) half-integer (with appropriate parity) 
comprised in the inverval $[Z_s(-\infty)/2\pi,Z_s(+\infty)/2\pi]$. 
This definition and \Zinf\ imply that
\eqn\Imax{\eqalign{
I^{max}_s+{1\over 2} &= {1\over 2} (M_s+r_s)
-E\left[{1\over 2}+{\gamma\over 2\pi} r_s\right]\cr
&+ \sgn \Mwd{s} -\sum_{t|\neigh} \MCbd{t}\cr
}}
and similarly
\eqn\Imin{\eqalign{
I^{min}_s-{1\over 2} &= -{1\over 2} (M_s+r_s)
+E\left[{1\over 2}+{\gamma\over 2\pi} r_s\right]\cr
&-\sgn \Mwu{s} +\sum_{t|\neigh} \MCbu{t}\cr
}}
Note that $I^{max}_s$ and $I^{min}_s$ have the correct parity (opposite
of $M_s+r_s$). 

Now it is recalled that for half-integer values of $Z_s$ on
the real axis, we have real roots and holes 
(including special roots/holes). Using
the obvious relation $M_s=\MC{s}+\MR{s}$, we find that
\eqn\MHrel{\MH{s}=I^{max}_s-I^{m in}_s+1 - M_s +\MC{s} + 2 \MS{s}}
Combining \Imax, \Imin, \MHrel\ and \rep,
we finally have
\eqn\qn{\eqalign{
r_s &= \MH{s} - 2 \MS{s} + 2 E\left[{1\over 2}+{\gamma\over 2\pi} r_s\right]\cr
&-\Mc{s}-2\th \Mw{s} +\sum_{t|\neigh} \MCb{t}\cr
}}

Once we have obtained Eq. \qn, we can take the scaling limit in it. The only simplification
that occurs concerns the special roots/holes. The situation is identical to the one
encountered in \DDVexc, so we shall not justify in detail the following statement: as 
$\theta\rightarrow\infty$, some special roots/holes are sent to infinity, and their
number exactly cancels the $2 E\left[{1\over 2}+{\gamma\over 2\pi} r_s\right]$ in \qn.
Finally, still calling $\MS{s}$ the number of remaining special roots/holes, we have
\eqn\qnb{
r_s = \MH{s} - 2 \MS{s} 
-\Mc{s}-2\th \Mw{s} +\sum_{t|\neigh} \MCb{t}}
Alhough we shall not use this simplification in the subsequent
calculations,
it is in fact essential for their self-consistency.

\newsec{Large $L$ limit}
We shall only sketch the $L\rightarrow\infty$ limit, in which we
should recover the usual physics of the infinite volume system.
Starting from the NLIE. \dere\ one should be able to generalize to results
of \ref\BDVV{O.~Babelon, H.J.~De~Vega and C.M.~Viallet,
{\it Nucl.\ Phys.} B220 (1983), 283.}
to all regimes and all simply laced Lie algebras.

For all values of $\gamma$, the holes correspond to relativistic
physical excitations that we
identify with solitons. From \dere\ and \qn\
we infer that a hole of type $s$ with
rapidity $\eta_{s,k}$ corresponds to a soliton of mass $m_s$ and
which belongs to the fundamental representation $w_s$ 
of $U_q(\widehat{\goth g})$. For example for $\goth g=A_1$, solitons and antisolitons
are put together in a $U_q(\widehat{\goth{sl}(2)})$ doublet, so here the holes correspond to solitons.

For the interpretation of the complex roots,
we need to continue the NLIE \dere\
to the whole complex plane.
The continuation is easily accomplished
if one correctly takes into account the poles one
catches when deforming the integration paths; we shall not describe
explicitly the whole procedure since it
is perfectly similar to what has already been done (Eqs. \defX, \defe).
We shall formally write the result as
\eqn\NLIE{Z_s(\lambda)= {m_s L\over 2}  (e(\lambda)-e(-\lambda))
+\sum_{t=1}^n X_{st} \star Q_t
+ g_s(\lambda)}
where for $\lambda$ real,
\eqn\defg{g_s(\lambda)\equiv\sum_{t=1}^n\left[
\sum_{k=1}^{\MH{t}} \chi_{st}(\lambda-\eta_{t,k})
-2\sum_{k=1}^{\MS{t}} \chi_{st}(\lambda-\sigma_{t,k})
-\sum_{k=1}^{\MC{t}} \chi_{st}(\lambda-\xi_{t,k})\right].}

Then we impose the relation $\exp(iZ_s(\xi_{s,k}))=(-1)^{1+\delta_s}$
for all complex roots $\xi_{s,k}$. The divergent imaginary part
of $mL (e(\xi_{s,k})-e(-\xi_{s,k}))$ has to be compensated by a pole
in $\exp(g_s(\xi_{s,k}))$; this forces the complex roots to fall
into certain configurations.

In the repulsive case ($\gamma<\pi/2$), the close roots
group into {\it quartets} which consists of two roots of the first
kind $\xi$, $\bar\xi$ ($\Im\xi>0$) and two roots of the second
kind $\xi-2i\pi/\cox$, $\bar\xi+2i\pi/\cox$. This configuration can
degenerate into a {\it 2-string} $\xi$, $\bar\xi$, with
$\Im\xi=i\pi/\cox$. The contribution of the different members of the
quartet to the energy exactly cancels, so that quartets have zero
energy. Wide roots do not have any constraints on their rapidities
since $e(\xi)=0$ for a wide root; for the same reason they do not
contribute to the energy.

The interpretation of this result is that complex roots serve as a way
of lowering the weight $r$ of the system without changing its energy.
Note that close roots and wide roots do not modify in the same
way $r$: we can rewrite \qnb\ using our knowledge of the configurations of close roots
\eqn\qnc{r_s=\MH{s}-2\MS{s} - {1\over 2} \sum_t C_{st} \Mc{t} -2 \Mw{s}}
The fact that the energy is unchanged
when adding complex roots in a system with fixed holes is the sign
of an enlarged symmetry at $L\rightarrow\infty$ (quantum {\it affine}
symmetry $U_q(\hat{\goth g})$).

The attractive regime is more complicated. Here we shall make some general observations.
Eqs. \ECb\ and \ECc\ show that wide roots now carry energy: their presence is related
to the appearance of {\it breathers}. By definition we call breathers all the particles
of the spectrum which are not the fundamental solitons. In the repulsive regime, the fundamental
solitons form no other bound states than themselves; but in the attractive regime, new bound states
are created. According to \qnb, breathers
are necessarily neutral for $\pi/2<\gamma<2\pi/3$, whereas they can be charged for $\gamma>2\pi/3$.
A more detailed description of the allowed configurations will be given in a coming paper.

Finally, interpreting the NLIE as equations for phaseshifts of physical particles on the
periodic space of length $L$, one immediately identifies $\chi_{st}(\lambda)$ with the phaseshift
between two solitons of type $s$ and $t$ with rapidity difference $\lambda$. More precisely,
this corresponds to scattering in the highest weight in the tensor product (insertion of
complex roots allows to obtain the lower weights). For example,
for $\goth g=A_n$, one has (using the expression for $X_{11}(\kappa)$ given in \XFour)
\eqn\Soneone{S_{11}(\lambda)=\exp\left(i \int_0^{+\infty} d\kappa\, 
{2\sin(\kappa\, h\lambda/\pi)\over \kappa}
{\sinh((\pi/\gamma-h)\kappa) \sinh(\kappa)\over \sinh((\pi/\gamma-1)\kappa) \sinh(h\kappa)}
\right)}
(up to a global phase). For $n=1$ this reproduces
the well-known Sine--Gordon soliton-soliton $S$-matrix.
For $n>1$ it is precisely
the $S$-matrix conjectured in \ref\HOL{T.J.~Hollowood,
{\it Int.\ J.\ Mod.\ Phys.} A8 (1993), 947.} for affine Toda with imaginary coupling.

\newsec{Large $\theta$ limit (decoupling of the 2 chiralities)}
In preparation for the UV (conformal) limit $L\rightarrow 0$, we shall first consider
the limit $\theta\rightarrow\infty$, with $M$ large but finite. Intuitively,
since $mL=M\e{-\theta}$, this is basically the same as the limit $L\rightarrow 0$.
Indeed, one can check that there is proper commutation of the limits, so that the results
we shall obtain in this section will be valid in the next one, in which we take
$L\rightarrow 0$ {\it after} the scaling limit $M\rightarrow\infty$, $\theta\rightarrow\infty$.
The advantage of keeping $M$ finite is that just as in section 5,
one can write intermediate equations which would diverge as $M\rightarrow\infty$.

In the large $\theta$ limit, the NLIE (just like the TBA
equations)
exhibit decoupling of the two chiralities. The functions
$Z_s(\lambda)$ have a growing flat plateau
in the region $[-\theta,\theta]$, which implies that
if we consider positions of roots and holes varying continuously with $\theta$,
then the set of roots and holes
divides into left-movers and right-movers, according to:
\eqn\mov{\eqalign{
\lambda_{s,k}&=\lambda_{s,k}^\pm \pm \theta\cr
\eta_{s,k}&=\eta_{s,k}^\pm \pm \theta\cr
}
}
where the $\lambda_{s,k}^\pm$ and $\eta_{s,k}^\pm$
($k=1\ldots M_s^\pm,\MH{s}^\pm$ after reordering
of the indices) are kept
fixed as $\theta\rightarrow\infty$. In particular for complex roots
and special roots/holes we define the 
$\xi_{s,k}^\pm$ and $\sigma_{s,k}^\pm$ ($k=1\ldots\MC{s}^\pm, \MS{s}^\pm$).
We allow exceptional unmoving roots or holes which may appear in special
configurations,
even though they will play no role for us; all it means is that
we do not impose for the moment relations like $\MH{s}^+ +\MH{s}^-=\MH{s}$.
We also define for future use $r_s^\pm\equiv
M\delta_{s1}-\sum_{t=1}^n C_{st} M_t^\pm$, in analogy with the
corresponding expression for $r_s$.

The next step is to define the two decoupled counting functions
$Z_s^{\pm}$ as
\eqn\Zpm{Z_s^{\pm}(\lambda)=\lim_{\theta\rightarrow +\infty} 
Z_s(\lambda\pm\theta)}
In the intermediate region, $Z_s(\lambda)$ is flat, 
so that $Z_s^+(-\infty)=Z_s^-(+\infty)$,
except if there are unmoving roots. To see this more clearly,
let us pick $Z_s^+(-\infty)$, and compute it $\mod 2\pi$. 
The relations we find
will be useful in the calculation of the UV conformal weights.

From the definition \defZ\ of $Z_s$, separating right-movers from the other
roots, we have
\eqn\om{\eqalign{
Z_s^+(-\infty) &= -  (\pi-2 \gamma) (M_s-2M_s^+)
+ \sum_{t|\neigh} (\pi-\gamma) (M_t-2M_t^+) \cr
&+ 2\pi \,\sgn (\Mwd{s}-\Mw{s}^+)
- 2\pi \sum_{t|\neigh} (\MCbd{t}-\MCb{t}^+)\cr
&= (\pi-\gamma) (r_s - 2r_s^+) + \pi (M_s - 2 M_s^+)\cr
&+ 2\pi \,\sgn (\Mwd{s}-\Mw{s}^+)
- 2\pi \sum_{t|\neigh} (\MCbd{t}-\MCb{t}^+)\cr
}}

We now introduce $z_s^+ \ \equiv Q_s^+(-\infty)$, so that
\eqn\omb{
\eqalign{
z_s^+ &= Z_s^+(-\infty) + \pi \delta_s \mod 2\pi\cr
&= Z_s^+(-\infty) - 2\pi \left(I^{min}_s{}^+ - {1\over 2}\right)\cr
&= Z_s^+(-\infty) - 2\pi \left( I^{max}_s + {1\over 2}
- (\MH{s}^+ -2\MS{s}^+ + M_s^+ - \MC{s}^+) \right)\cr
&= -2(\pi-\gamma)r_s^+ + 2\pi(\MH{s}^+-2\MS{s}^+)
-\left( \gamma r_s - 2\pi E\left[ {1\over 2} + {\gamma\over 2\pi} r_s \right]
\right)\cr
&-2\pi (\Mc{s}^+ +2\th \Mw{s}^+)  
+ 2\pi \sum_{t|\neigh} \MCb{t}^+\cr
}
}
Here, we have introduced $I_s^{min}{}^+$, the smallest half-integer (with
appropriate parity) larger than $Z_s^+(-\infty)$, and related it to
$I_s^{max}$ in the obvious way; and we have replaced $Z_s^+(-\infty)$ with
its value \om.

Therefore we are led to the particularly simple expression
\eqn\omfin{z_s^+ = \gamma (2r_s^+ - r_s) + 2\pi (\hat{r}_s^+ -r_s^+)}
with
\eqn\rhat{\eqalign{
\hat{r}_s^+ &\equiv \MH{s}^+ - 2 \MS{s}^+
+ E\left[{1\over 2}+{\gamma\over 2\pi}
r_s \right]\cr
&-\Mc{s}^+ -2\th \Mw{s}^+ + \sum_{t|\neigh} \MCb{t}^+\cr
}}
Comparing \omfin\ with \qn, one can intepret $\hat{r}_s^+$ as the
partial quantum number induced by the right-movers.
Of course, in the scaling limit, the term $E\left[{1\over 2}+{\gamma\over 2\pi}
r_s \right]$ is cancelled by the extremal special roots/holes, and
we can simply remove it in \rhat\ (cf \qnb).

A similar expression may be found for $z_s^-\equiv Q_s^-(+\infty)$:
\eqn\omfinb{z_s^-=-\gamma(2r_s^- - r_s) - 2\pi (\hat{r}_s^- -r_s^-)}
In general, \omfin\ and \omfinb\ do not coincide. However, if we suppose that
we are in a generic situation so that there are no unmoving roots, then $r_s=r_s^+ + r_s^-$
and defining $\Delta r_s = r_s^+ - r_s^-$ we find
\eqn\omfinc{z_s^\pm=\gamma \Delta r_s - 2\pi E\left[ {1\over 2} + {\gamma\over 2\pi} \Delta r_s
\right]}

\newsec{Computation of the UV conformal weights.}
We shall now use the powerful machinery of the NLIE to probe the
physics of the UV region of our model. Indeed, it is expected that
as $L\rightarrow 0$ (the same as $T\rightarrow\infty$ in 
TBA equations), the theory should flow to its UV fixed
point. More precisely, the leading $1/L$ behavior of the energy of
the excited states should coincide with the results of CFT,
giving us an explicit expression of the central charge and all
conformal weights.

According to the remarks made at the beginning of previous section, all
the results obtained in it are valid if we first send $M$ and $\theta$ to
infinity so that $mL$ remains finite, then consider the limit
$mL\rightarrow 0$. In particular we define again left/right-movers: 
($r\equiv mL$)
\eqn\movb{\eqalign{
\lambda_{s,k}&=\lambda_{s,k}^\pm \pm \log(2/r)\cr
\eta_{s,k}&=\eta_{s,k}^\pm \pm \log(2/r)\cr
}
}
and the chiral counting functions:
\eqn\Zpmb{Z_s^{\pm}(\lambda)=\lim_{r\rightarrow 0} 
Z_s(\lambda\pm\log(2/r))}

In the chiral limit the NLIE \NLIE\ becomes
\eqn\NLIEch{
Z^\pm_s(\lambda)=\pm{m_s\over m} e(\pm\lambda)
+\sum_{t=1}^n X_{st} \star Q_t^\pm
+g_s^\pm(\lambda)}
where $Q^\pm$ (resp. $g_s^\pm$)
is related to $Q$ (resp. $g_s$) in the obvious way.

Now we begin the computation of the finite-size corrections to
the energy. We recall that
\eqn\Erec{E=E^+ + E^-}
with $E^\pm=(E\pm P)/2$.
Let us choose $E^+$; we expand it
from \Escal\ and keep the dominant term in the
$L\rightarrow 0$ limit:
\eqn\Edera{
E^+= {1\over L} \sum_s {m_s\over m}
\left[ \sum_{k=1}^{\MH{s}^+} e(\eta_{s,k}^+)
-2\sum_{k=1}^{\MS{s}^+} e(\sigma_{s,k}^+)
- \sum_{k=1}^{\MC{s}^+} e(\xi_{s,k}^+)
-{1\over 2\pi} \int d\lambda \, \e\lambda \, Q_s^+(\lambda) \right]
}
We have used the notation $e(\lambda)$ even for holes and special
roots/holes for which one has of course $e(\lambda)=\e\lambda$.
We now use the NLIE \NLIEch\ to get rid of the $e(\lambda)$
terms: since $Z_s^+(\eta_{s,k}^+)=2\pi \IH{s,k}^+$ and similar
relations for special roots/holes and complex roots, we find that
\eqn\Ederb{
\eqalign{
E^+= {1\over L} \Bigg[ &2\pi(I_H^+ -2I_S^+ -I_C^+)
+\sum_s \Bigg[ -\sum_{k=1}^{\MH{s}^+} g_s^+(\eta_{s,k}^+)
+2\sum_{k=1}^{\MS{s}^+} g_s^+(\sigma_{s,k}^+)
+\sum_{k=1}^{\MC{s}^+} g_s^+(\xi_{s,k}^+) \cr
& - {1\over 2\pi} \int d\lambda \, (d/d\lambda) f_s^+(\lambda) \,
Q_s^+(\lambda) \Bigg]\Bigg] \cr
}
}
We have introduced the notation
$f_s^+(\lambda)\equiv {m_s\over m} \e\lambda + g_s^+(\lambda)$
to recombine the differents terms where $Q_s^+$ appears.
$I_H^+\equiv\sum_s \sum_k \IH{s,k}$, $I_S^+\equiv\sum_s \sum_k \IS{s,k}$,
$I_C^+\equiv\sum_s \sum_k \IC{s,k}$.

Next we use a variant of the dilogarithm trick: it is the
multi-component generalization of the lemma of \DDVexc. We state the equality:
\eqn\lem{\eqalign{
\sum_s \int d\lambda \, (d/d\lambda) &f_s^+(\lambda) \, Q_s^+(\lambda)
= - 2 \sum_s \Re \int_{\Gamma_s} {du\over u} \log(1+u)\cr
& -{1\over 2}\sum_{s,t} \left[ Q_s^+(+\infty)Q_t^+(+\infty)-Q_s^+(-\infty)Q_t^+(-\infty) \right]
\int_{-\infty}^{+\infty} dx\, X_{st}(x)
}}
where $\Gamma_s$ is a contour in the complex plane which goes from
$(-1)^{\delta_s}Z_s^+(-\infty+i0)$ to $(-1)^{\delta_s}
Z_s^+(+\infty+i0)$ avoiding the logarithmic cut on $[-\infty,-1]$.

Using $\int_{-\infty}^{+\infty} dx\, X_{st}(x)=X_{st}(\kappa=0)={1\over\pi}\chi_{st}(+\infty)
=\delta_{st}-C^{-1}_{st}/(1-\gamma/\pi)$ and computing explicitly the integral over $u$, we find:
\eqn\Ederc{
\eqalign{
\sum_s \int d\lambda \, (d/d\lambda)  f_s^+(\lambda) \, Q_s^+(\lambda)
&=\sum_s \left( {\pi^2\over 6} - {z_s^+{}^2 \over 2}\right)
+{1\over 2} \sum_{s,t} z_s^+ z_t^+ X_{st}(k=0)\cr
&=n {\pi^2\over 6} - {1\over 2} \sum_{s,t} z_s^+ z_t^+
{C_{st}^{-1}\over 1-\gamma/\pi}\cr
}
}
Note that no dilogarithm function is actually involved, only
elementary functions appear.

The sum of all $g_s^+$ appearing in \Ederb\ simplifies enormously due to the oddness under
simultaneous exchange of $s\leftrightarrow t$ and $\lambda\leftrightarrow -\lambda$
in $\chi_{st}(\lambda)$; after some lengthy algebra we find that
\eqn\sumg{
\sum_s \Bigg[ -\sum_{k=1}^{\MH{s}^+} g_s^+(\eta_{s,k}^+)
+2\sum_{k=1}^{\MS{s}^+} g_s^+(\sigma_{s,k}^+)
+\sum_{k=1}^{\MC{s}^+} g_s^+(\xi_{s,k}^+) \Bigg]
= - \sum_{s,t} \chi_{st}(+\infty) \hat{r}_s^+ (r_t - \hat{r}_t^+) + 2\pi q^+
}
where $q^+$ is a half-integer which depends on the number of complex roots (wide roots,
roots of the first kind).

Putting everything together, the energy takes the form
\eqn\Ederd{
\eqalign{
E^+&={1\over L} \left[ -n {\pi\over 12} 
+2\pi(I_H^+ -2I_S^+ -I_C^+ +q^+)\right. \cr
&\left. + \sum_{s,t} z_s^+ z_t^+
{C_{st}^{-1}\over 4\pi(1-\gamma/\pi)} 
-\pi \sum_{s,t} \hat{r}_s^+ (r_t-\hat{r}_t^+) 
\left( \delta_{st}-{C_{st}^{-1}\over 1-\gamma/\pi} \right)
\right]\cr}
}
Using the expression \omfin\ for $z_s^+$ and performing some
recombinations, we can write the final result
\eqn\Edere{
E^\pm={2\pi\over L} \left( -{c\over 24} + \Delta^\pm + p^\pm \right)
}
where
\eqn\cc{c=n}
is the central charge,
\eqn\cw{
\Delta^\pm={\sum_{s,t} C_{st}^{-1} 
\left[ r_s + (1-\gamma/\pi) (2r_s^\pm-r_s) \right]
\left[ r_t + (1-\gamma/\pi) (2r_t^\pm-r_s) \right]
\over 8(1-\gamma/\pi)}
}
are the (primary) conformal weights, and
\eqn\cdesc{
p^\pm=\pm(I_H^+-2I_S^+-I_C^+ +q^+)
-{1\over 2}\sum_s  \hat{r}_s^\pm
(r_s-\hat{r}_s^\pm +M_s - 2M_s^\pm)
}
is a half-integer. In view of \Edere\ one can reasonably assume
that $p^\pm$ is in fact
non-negative, which can be checked directly.

We now exclude special configurations with unmoving roots, so that
$r_s=r_s^+ + r_s^-$, and $2r_s^\pm - r_s=\pm \Delta r_s$
with, as before, $\Delta r_s\equiv r_s^+ - r_s^-$\foot{Special
attention must be paid to the case $\Delta r_s=0$, in
which, to avoid an unmoving root at $Z=0$ (cf Eq. \omfinc), one
must choose the appropriate value of $M_s \mod 2h$ so
that $\delta_s=0$.}. This slightly simplifies the
form of \cw, and allows the following interpretation:
the central charge \cc\ indicates $n$ free bosons. In fact general
arguments (see appendix A) show that the UV fixed point of affine
Toda with imaginary coupling should be a multi-component Coulomb gaz
(i.e.\ compactified free bosons). Indeed,
the conformal weights \cw\ are closely related
to those of $n$ free bosons, as is shown in appendix A. For example,
in the $A_1$ case, they are connected with the deformed chiral
Gross--Neveu model, whose bosonization is the Sine--Gordon model.
The ultra-violet
conformal weights \cw\ are related to the infra-red
conformal weights of the corresponding spin chain \ref\DVSPT{H.J.~De~Vega,
{\it J.\ Phys.} A21 (1988), L1089\semi
J.~Suzuki, {\it J.\ Phys.} A21 (1988), L1175\semi
A.G.~Izergin, V.E.~Korepin and N.Yu.~Reshetikhin,
{\it J. Phys.} A22 (1989), 2615\semi
S.V.~Pokrovsky and A.M.~Tsvelick, {\it Nucl.\ Phys.} B320 (1989), 696.} by 
exchange of $r_s$ and $\Delta r_s$.

It should be pointed out that the finite-size correction to the energy 
{\it does not depend} on the actual values of the rapidities of holes and complex roots:
it only depends on their number, or more precisely of the partial (chiral) 
$\goth g_0$ quantum numbers.
In particular this indicates that the string hypothesis, which constrains the positions
of the complex roots, is useless here. Indeed we have not made any use of it, knowing
that for low-lying excited states it is in fact violated.

\newsec{Twist and quantum group truncation}
In the Sine--Gordon model, it is known that at rational values
of $\gamma/\pi$, one can consistently restrict the theory to a smaller Hilbert
\nref\SMI{F.A.~Smirnoff, {\it Nucl. Phys.} B337 (1990),
156.}\nref\LBL{A.~Leclair, {\it Phys. Lett.} B230 (1989), 103\semi
D.~Bernard and A.~Leclair, {\it Nucl. Phys.} B340 (1990),
72.}[\xref\SMI,\xref\LBL] which in particular displays a different UV
behavior, reproducing the minimal models. We shall show that such
a truncation can be extended to the complex affine Toda model.

The key ingredient of the truncation is the quantum group symmetry
and its representation theory \nref\PAS{V.~Pasquier, {\it Nucl. Phys.}
B295 (1988), 491.}[\xref\PAS,\xref\LBL]. Since the representation
theory of $U_q(\goth g)$ is well-developed and resembles closely that
of $U_q(\goth{sl}(2))$, we expect no particular difficulty. However,
implementing the truncation in the Bethe Ansatz framework raises
several questions.

The natural way to implement the truncation is to introduce a twist in the B.A.E.
Indeed it is known that B.A.E. with twist
\ref\DVG{H.J.~De~Vega and H.J.~Giacomini, {\it J.\ Phys.} A22 (1989), 2759.}
are related to RSOS models, which themselves are equivalent to
restricted Sine--Gordon (at least in the UV limit),
but this is a rather indirect connection, and
we would like to have a more direct derivation of the truncation.
The second problem is specific to the NLIE approach: as we are
considering the theory on a compactified space of length $L$, it {\it
does not} possess the quantum group $U_q(\goth g)$ symmetry. To
summarize, even in the $U_q(\goth{sl}(2))$ case, in which the
twist in the DDV equations (NLIE) has been done
\ref\FMQR{D.~Fioravanti, A.~Mariottini,
E.~Quattrini and F.~Ravanini, {\it Phys.\ Lett.} B390 (1997), 243.},
it has not been justified that this procedure was the same as the
quantum group truncation discussed earlier. We shall give now such
a justification for the generalized case of affine Toda. The quantum
group symmetry will reappear after a modular transformation which we are
naturally led to doing.
In the UV limit we shall find results which bear the same connection
to the JMO models \ref\JMO{M.~Jimbo, T.~Miwa and M.~Okado, {\it
Mod.\ Phys.\ Lett.} B1 (1987), 73.} as restricted
Sine--Gordon to the RSOS models.

\subsec{The group-theoretic background.}
Let us remind the reader that the affine Toda model (with imaginary coupling constant)
associated to the simply laced Lie algebra $\goth g$
consists of $n$ bosonic fields, grouped into a field $\phi$ which 
belongs to the Cartan subalgebra $\goth g_0$.
The action is given by
\eqn\lagToda{{\cal S}={1\over\beta^2}\int d^2 x \left[
(\der_\mu \phi)^2 + m^2 \sum_{s=0}^n
\exp\left(-i \left< \alpha_s,\phi\right>\right)\right]
}
The $\alpha_s$, $s=0\ldots n$ are the simple roots of
$\hat{\goth g}$ (alternatively one can consider that 
the $\alpha_s$, $s=1\ldots n$ are the simple roots of $\goth g$,
and $-\alpha_0$ is the highest root of $\goth g$).

Since $\goth g_0$ possesses a scalar product, we identify it with its dual space (weight space).
With this convention, one can decompose $\phi$ in the basis of fundamental weights:
$\phi=\sum_s \phi_s w_s$. Using the orthogonality relations
$\left<\alpha_s,w_t\right>=\delta_{st}$
($s,t=1\ldots n$), it is obvious to check that
this model has a $\Bbb Z^n$ symmetry, with generators
$T_s: \phi_s \to \phi_s+2\pi$. In order to
select the eigenvalues of the $T_s$, one
introduces the ``shifted'' partition function\foot{A more standard
denomination would be ``twisted'' partition function, but we reserve the word
``twisted'' for a slightly different -- in fact, dual -- situation, cf \resZ.}
$\Z_k$: ($k=(k_1,\ldots,k_n)\in {\Bbb Z}^n$)
\eqn\shiZ{\eqalign{
\Z_k&\equiv \tr(\exp(-\beta H_L)T_1^{k_1}\ldots T_n^{k_n})\cr
&= \int_{\scriptstyle\phi(t=\beta,x)=\phi(t=0,x)+2\pi k
\atop\scriptstyle \phi(t,x=L)=\phi(t,x=0)\, \mod{2\pi}} \,[d\phi]\,
\e{-{\cal S}[\phi]}
}}
We have considered the model on a
finite space of size $L$ and have imposed periodic boundary conditions
{\it modulo $2\pi$ only} for the $\phi_s$. $H_L$ is
the Hamiltonian in the corresponding operator formalism.
We have also taken a finite-temperature $\beta$ (this $\beta=1/T$ has nothing to do,
of course, with the constant in front of the action \lagToda); later, when we are
concerned with
the ground state and low-lying excited states only, we shall take
the limit $\beta\rightarrow\infty$, these states corresponding to
the first terms in the large $\beta$ expansion.

Next we introduce the partition function restricted
to the sector of the Hilbert
space of Toda, in which the $T_s$ have the eigenvalues 
$\e{i\omega_s}$:
\eqn\resZ{\eqalign{
\Z(\Omega)&\equiv \tr_\Omega(\exp(-\beta H_L))\cr
&=\sum_{k_1,\ldots,k_n=-\infty}^{+\infty}
\e{i\sum_{s=1}^n \omega_s k_s} \Z_k
}}

The eigenvalues are parametrized by
$\Omega\in \exp(i\goth g_0)$: $\Omega=\exp(i\omega)$ where $\omega
=(\omega_1,\ldots,\omega_n)$ in the basis of fundamental weights (after identifying,
as above, $\goth g_0$ and weight space).
$\tr_\Omega$ means trace in the sector $T_s=\e{i\omega_s}$.

In order to understand why these subtleties are usually neglected,
let us first consider the $L\rightarrow\infty$ (infinite space)
limit: then the transition (in time) between different classical vacua
is suppressed and $Z_k\rightarrow 0$ for $k\ne 0$; $\Z(\Omega)$ becomes
independent of $\Omega$ i.e.\ all the sectors of Toda become degenerate.

The functional integral \resZ\ can receive another interpretation
by exchanging the roles of space and time; after this
modular transformation, the operatorial interpretation becomes
\eqn\ferZc{
\Z(\Omega)= \tr_1(\exp(-L H_\beta) \Omega),}
($\tr_1$ means trace over the trivial sector of the ${\Bbb Z}^n$ symmetry, 
i.e.\ $\phi_s\equiv \phi_s+2\pi$)
with $\Omega$ considered as the exponential
of an element of the Cartan algebra 
$\goth{g}_0$ (acting on the whole Hilbert space).
This formula can
be guessed by noticing that after the modular
transformation, the numbers $k_s$ describe precisely the topological
charges which are associated with the $\goth{g}_0$ symmetry. We shall
call $\Z(\Omega)$ the twisted partition function since both in the
transfer matrix language (see next paragraph) or in a ``fermionized''
language (using boson-fermion equivalence in 2D;
though this introduces additional subtleties due to
fermionic boundary conditions and modular invariance that we do not
wish to discuss) $\Omega$ appears as a twist in the spatial boundary conditions.

Let us now consider the limit $\beta\rightarrow\infty$.
In this limit, 
$H_\beta$ should commute with the action of the
full quantum group $U_q(\goth{g})$, enlarging the $\goth{g}_0$
symmetry. Then one can decompose the Hilbert space according
to $U_q(\goth{g})$ representations, and use a 
character expansion\foot{Note that decomposition
\chrZ\ is not the same decomposition as \resZ: in \chrZ\ the sum
is over all {\it highest} weights of $U_q(\goth g)$ whereas in \resZ\ it is
over all (integral) weights.}:
\eqn\chrZ{\Z(\Omega)=\sum_R \chi_R(\Omega) \Z_R}
where $R$ runs over all highest weight representations of
$U_q(\goth{g})$, and $\Z_R$ is the partition function of the
sector of the Hilbert space with representation $R$ (divided by
the dimension of the representation).

So far, in all the previous sections of this paper
we have implicitly chosen $\Omega=1$:
this also corresponds, from what has been said,
to the ``trivial'' sector of Toda for the $\Bbb Z^n$ symmetry.
We shall now choose a non-trivial $\Omega$ which selects 
``good'' representations of $U_q(\goth{g})$ for $\gamma/\pi$
rational. More precisely, we choose
\eqn\chOm{
\Omega=q^{2H}\equiv q^{\sum_{\alpha>0} H_\alpha}
}
where $H_\alpha$ is the element of the Cartan algebra $\goth g_0$ associated to the positive
root $\alpha$. This corresponds more explicitly to
$\omega_s=2\gamma$\footnote{$^\ddagger$}{The factor of $2$ originally comes from
our convention for the definition of the deformation parameter
$q$; other authors use $q'\equiv q^2$, which removes this $2$.}.
Then it is known that $\chi_R(\Omega)=0$ for all the ``bad'' representations
(indecomposable but not irreducible representations, and a few others),
and, according to \chrZ, we are left with contributions from the
``good'' representations, with
prefactors $\chi_R(\Omega)$ which correctly account
for the truncation of the tensor product (for a more
thorough analysis in the $A_1$ case see [\xref\LBL,\xref\PAS]).

It is worth stressing that {\it any} value of $\Omega$ is a priori
conceivable: the spectrum of the generators $T_s$ is the whole $U(1)$
circle (contrary to what has been written in the recent literature).
This does not contradict the quantum group truncation, because
one should be careful that the truncation takes place when the {\it
time} direction is compactified with length $L$, i.e.\ after a modular
transformation has been done. In particular the ``ground state''
contribution for twisted Toda does not correspond at all to the ground state
contribution in this dual picture: on the contrary we are considering
the theory at finite temperature $T=1/L$ (which we
eventually send to infinity when we look at the UV region). So the
finite-size correction varies continuously with the twist $\Omega$,
as we shall see in next paragraph, but only for particular (discrete)
values does it have an interpretation in terms of a truncated Hilbert space.

\subsec{Twist and Bethe Ansatz.}
It is particularly simple to add a twist in our formalism: the
twisted version of transfer matrix \Tmat\ is
\eqn\Ttwi{T(\Lambda,\Theta,\Omega)=\tr_{aux}\left[
L_1(\Lambda-i\Theta) L_2(\Lambda+i\Theta) \ldots L_{2M-1}(\Lambda-i\Theta)
L_{2M}(\Lambda+i\Theta)\Omega\right]}
$\Omega$ acts in the auxiliary space.
In the scaling limit, one can easily convince oneself that
the twisted transfer matrix
leads to the model described by partition function $\Z(\Omega)$ of \resZ.

Of course, the addition of the twist preserves the integrability;
to diagonalize $T$ we now have twisted Bethe Ansatz Equations:
\eqn\NBAE{\eqalign{
&\prod_{j=1}^{M_s}
{\sinh(\gamma(
{\cox\over 2\pi}(\lambda_{s,k}-\lambda_{s,j})+i))
\over\sinh(\gamma({\cox\over 2\pi}(\lambda_{s,k}-\lambda_{s,j})-i))}
\prod_{t|\neigh} \prod_{j=1}^{M_t}
{\sinh(\gamma(
{\cox\over 2\pi}(\lambda_{s,k}-\lambda_{t,j})-i/2))
\over\sinh(\gamma({\cox\over 2\pi}(\lambda_{s,k}-\lambda_{s,j})+i/2))}
\cr
&=\e{i\omega_s} \left[{\sinh(\gamma(
{\cox\over 2\pi}(\lambda_{s,k}-\theta)+i/2))
\over\sinh(\gamma({\cox\over 2\pi}(\lambda_{s,k}-\theta)-i/2))}
{\sinh(\gamma({\cox\over 2\pi}(\lambda_{s,k}+\theta)+i/2))
\over\sinh(\gamma({\cox\over 2\pi}(\lambda_{s,k}+\theta)-i/2))}
\right]^{M\delta_{s1}}
\cr}
}

Finally, this introduces an extra term $\omega_s$
in the definition of the counting function $Z_s$, and
the NLIE \dere\ becomes:
\eqn\deretwi{\eqalign{
Z_s= m_s L \sinh \lambda
+\sum_{t=1}^n \Bigg[ &{C_{st}^{-1}\over 1-\gamma/\pi} \omega_t
+X_{st} \star Q_t
+\sum_{k=1}^{\MH{t}} \chi_{st}(\lambda-\eta_{t,k})\cr
&-2\sum_{k=1}^{\MS{t}} \chi_{st}(\lambda-\sigma_{t,k})
-\sum_{k=1}^{\MC{t}} \chi_{st}(\lambda-\xi_{t,k}) \Bigg]}
}

\subsec{The UV limit of the truncated theory.}
One can again probe the UV fixed point (of the truncated theory) by sending $L$ to $0$.
This amounts to redoing the calculations of section 8 in the presence
of the twist. 
We shall only rewrite the relations that are modified
in the process. Eq. \omfin\ becomes
\eqn\omfintwi{z_s^+=\gamma(2r_s^+ -r_s)+2\pi(\hat{r}_s^+ - r_s^+) + \omega_s} 
When going from \Edera\ to \Ederb\ one uses the NLIE, so one gets
an extra term:
\eqn\Ederbtwi{
\eqalign{
E^+= {1\over L} \Bigg[ &2\pi(I_H^+ -2I_S^+ -I_C^+)
+\sum_s \Bigg[ -\sum_{k=1}^{\MH{s}^+} g_s^+(\eta_{s,k}^+)
+2\sum_{k=1}^{\MS{s}^+} g_s^+(\sigma_{s,k}^+)
+\sum_{k=1}^{\MC{s}^+} g_s^+(\xi_{s,k}^+) \cr
& -\hat{r}_s^+ \sum_t {C^{-1}_{st}\over 1-\gamma/\pi} \omega_t
- {1\over 2\pi} \int d\lambda \, (d/d\lambda) f_s^+(\lambda) \,
Q_s^+(\lambda) \Bigg]\Bigg] \cr
}
}
Finally $E^\pm$ is given by
\eqn\Ederdtwi{\eqalign{
E^\pm&={2\pi\over L} \Bigg[-{n\over 24} + p^\pm\cr
&+{\sum_{s,t} C_{st}^{-1}
\big[ r_s \pm (1-\gamma/\pi)\Delta r_s \mp\omega_s/\pi \big]
\left[ r_t \pm (1-\gamma/\pi)\Delta r_t \mp\omega_t/\pi \right]
\over 8(1-\gamma/\pi)}
\Bigg]}
}
where $p^\pm$ is unchanged. Note that this expression, just like
\cw, correctly behaves
under space parity: $E^+$ and $E^-$ (or $\Delta^+$ and $\Delta^-$)
are exchanged by
$r_s^\pm\leftrightarrow -r_s^\mp$ (the $\omega_s$, from their definition,
are unaffected by space parity).

For generic values of $\gamma$ and of the $\omega_s$ this formula simply gives
the finite-size corrections of affine Toda in a sector
$T_s=\e{i\omega_s}$. In particular, one finds that the true ground
state of the theory is in the sector $\Omega=1$, since for $r_s=0$
the energy increases as $\Omega$ moves away from $1$.

As explained in previous paragraph, the result \Ederdtwi\ acquires a new
significance for $\gamma/\pi$ rational and $\Omega$ fixed
by \chOm. The new central charge of the truncated theory
is smaller than $n$, since the second line
of \Ederdtwi\ is no longer purely quadratic in the $r_s$, $r_s^\pm$
(it has a constant and a linear part). 

Let us first consider $\gamma=\pi/(p+1)$. Setting
$\omega_s=2\gamma$, one finds the result (using the strange formula
$12\sum_{s,t} C_{st}^{-1}=h\, \dim\goth g$)
\eqn\cctwi{\left\{ \eqalign{
c&=n\left(1-{h(h+1)\over p(p+1)}\right)\cr
\Delta^\pm&=\Delta_0+
{\sum_{s,t} C_{st}^{-1}
\left[ m_s (p+1) \pm n_s p \mp 1 \right]
\left[ m_t (p+1) \pm n_t p \mp 1 \right]
\over 2p(p+1)}
}\right.}
where $\Delta_0\equiv (c-n)/24=-{n\over 24}{h(h+1)\over p(p+1)}$, and
$r_s=2m_s$, $\Delta r_s=2n_s$.
\cctwi\ is characteristic 
of a representation of the $W(\goth g)$ extended conformal algebra corresponding
to unitary RCFTs \ref\BG{A.~Bilal and J-L.~Gervais, {\it Nucl.\ Phys.} B318 (1989), 579.}.
For $\goth g=A_1$ and $n_1=0$
\cctwi\ is equivalent to what was found in \FMQR.

Let us now suppose that $\gamma=\pi (q-p)/q$ ($p$ and $q$ coprime integers).
Upon replacement of $\gamma$
and $\omega_s=2\gamma$ with their values one finds
\eqn\cctwib{
\left\{ \eqalign{
c&=n\left(1-h(h+1){(p-q)^2\over pq}\right)\cr
\Delta^\pm&=\Delta_0+
{\sum_{s,t} C_{st}^{-1}
\left[ m_s q \pm n_s p \mp (q-p) \right]
\left[ m_t q \pm n_t p \mp (q-p) \right]
\over 2pq}
}\right.}
with similar notations as in \cctwi;
$\Delta_0=-{n\over 24}{h(h+1)(p-q)^2\over pq}$.
This time we find the representations of $W(\goth g)$ corresponding to
all RCFTs $(p,q)$.

\newsec{Conclusion and prospects}
We have presented here some results concerning the affine Toda model associated
to a simply laced Lie algebra. We have written Non-Linear Integral Equations which
allow to interpolate excited states from $L=\infty$ (IR region) to $L=0$ (UV region).
The two limits have been discussed. In the UV region we recover results of CFT.
One should study more thoroughly the
$L\rightarrow\infty$ limit in the attractive regime:
it would give the full mass spectrum and scattering 
of the theory. This promises to be a rather complex task, because of the problem
of classifying the breathers.

Finally, the quantum group truncation has been described in detail, and the corresponding
NLIE written. We have checked that the
truncated theory does display a central charge and conformal weights which
are compatible with $W(\goth g)$ symmetry. However, this requires some further clarification:
indeed it is not completely obvious which primary operators are present, and under the form
of which states. For example,
in the $\goth g=A_1$ case, this is probably related
to the subtle differences which exist
between the various ``equivalent'' formulations of the model (cf
appendix A).
A similar analysis is probably possible for a general algebra $\goth g$,
but it has not been done yet.

%
\bigbreak\bigskip\bigskip\centerline{{\bf Acknowledgements}}\nobreak
I would like to thank J-L.~Gervais, F.~Smirnov, J-B.~Zuber and especially
D.~Bernard and H.~De~Vega for useful discussions.

\appendix{A}{Multi-component Coulomb gaz.}
There are many equivalent ways of introducing the multi-component
generalization of the conformal Coulomb gaz. The most appropriate one for us
is to start from action \lagToda; in the UV limit it can be shown that the
mass term (after appropriate renormalization), for $\beta^2<8\pi$,
tends to zero. 
Rewriting the remainder of the action in terms of the rescaled
components $\Phi_s=R\phi_s$ with $R\equiv \sqrt{4\pi}/\beta$
results in
\eqn\lagCoul{{\cal S}^{\rm Coulomb}={1\over 4\pi}\int d^2 x \sum_{s,t=1}^n C_{st}^{-1}
(\der_\mu\Phi_s)(\der_\mu\Phi_t)}
(the normalization of the action is conventional; it fixes the radius of
compactification, which we have chosen for $n=1$ as in
\ref\POLCH{J.~Polchinski, Lectures in Les Houches Summer School
``Fluctuating Geometries in Statistical Mechanics and Field Theory'' (1994).}).
From the discussion of section 9.1 it should also be clear that in
the trivial sector of the ${\Bbb Z}^n$ symmetry, one should identify
$\Phi_s$ and $\Phi_s+2\pi R$ which means we are dealing with
compactified free bosons on circles of radius $R$ (but not $n$ independent compactified bosons).

There are additionnal symmetries arising in this UV limit: besides
the topological currents $\epsilon_{\mu\nu}\der_\nu\Phi_s$ associated to our usual
$\goth g_0=\goth u(1)^n$ symmetry,
we have the obvious currents
$\der_\mu\Phi_s$ associated to $\Phi_s\to\Phi_s+{\rm cst}$ (the corresponding
symmetry group is
$U(1)^n$ since $\Phi_s\equiv\Phi_s+2\pi R$).
We now have two sets of quantum numbers describing a state:
the $\goth g_0$ quantum numbers $m_s$ (also called winding numbers
or magnetic charges) and the $e_s$ (``target space'' momenta in the
string picture, or electric charges).
Standard arguments \ref\stri{K.S.~Narain,
M.H.~Sarmadi and E.~Witten, {\it Nucl.\ Phys.} B279 (1987), 369\semi
J.~Bagger, D.~Nemeschansky, N.~Seiberg and
S.~Yankielowicz, {\it Nucl. Phys.} B289 (1987), 53.}
allow to find the full spectrum. The primary (for the $U(1)^n \times U(1)^n$
Kac-Moody algebra)
conformal weights are given by
\eqn\cwcou{\Delta^\pm={1\over 4}
\sum_{s,t} C_{st}^{-1} \left(\sum_{s'} C_{ss'} e_{s'}/R \pm m_s R\right)
\left(\sum_{t'} C_{tt'} e_{t'}/R \pm m_t R\right)}
where the $m_s$ and the $e_s$ are the aforementioned quantum numbers (integers).

Note in particular that the purely electric operators $\e{i <\alpha,\phi>}$
have dimension
\eqn\cwelec{\Delta^\pm={1\over 4} {<\alpha,\alpha>\over R^2}}
so that for the perturbing operators of \lagToda\ we have
$\Delta^\pm=1/2R^2=\beta^2/8\pi$. They are relevant for $\beta^2<8\pi$, as expected.

Naively,
there are several ways of matching the conformal weights \cw\ and
\cwcou, 
due to the many partial dualities relating different radii of
compactification. One finds that the correct
relation to impose is $\gamma=\pi-\beta^2/8$, so that $R$ is given by
\eqn\radius{R={1\over\sqrt{2(1-\gamma/\pi)}}}
and the identifications are
\eqna\ident$$\eqalignno{
r_s&=m_s&\ident a\cr
\Delta r_s &=2\sum_{t=1}^n C_{st} e_t
&\ident b\cr}$$
where $\Delta r_s=r_s^+ - r_s^-=\sum_t C_{st} (M_t^- - M_t^+)$ so that
$e_s={1\over 2}(M_s^- - M_s^+)$.

The first identification \ident a was expected on general grounds.
Note that due to its definition \rep, the $r_s$ span only
a subset of the integer lattice. One first constraint is that all $r_s$ are positive; this is
due to the fact that we are only considering highest weight states.
If we also considered lower weight states (e.g. antisolitons and not just solitons
for Sine--Gordon), it is expected that we would recover negative values.
Furthermore, the $r_s$ are always in a sublattice: for example,
for $\goth g=A_n$, one easily finds that
the $r_s$ satisfy the constraint $\sum_s s\, r_s\equiv 2M \mod n+1$ (conservation of the number
of boxes of the Young tableau $\mod n+1$).
However, as is usual in the Bethe Ansatz,
when $M$ is sent to infinity one can consider
all values of $2M \mod n+1$ simultaneously
(possibly considering an odd number of sites),
so that one recovers all possible $r_s$.

Let us now discuss briefly the allowed values of $e_s$: it would
seem that the $e_s$ can be half-integers (in fact, extrapolating
\ident b to arbitrary values of $\Delta r_s$,
one would even find that $e_s\in {1\over 2h}{\Bbb Z}$). The
situation is particularly clear in the $\goth g=A_1$ case, in which
$m=r$ and $e={1\over 4}\Delta r$. The correct interpretation of
this non-integerness is that
the Bethe Ansatz model we are considering does not describe
Sine--Gordon, but really an equivalent model: the deformed
$SU(2)$ chiral Gross--Neveu model \ref\dGN{The deformed $SU(2)$ chiral
Gross--Neveu model is studied via the Bethe Ansatz in\hfil\break
G.I.~Japaridze,
A.A.~Nersesyan et P.B.~Wiegmann, {\it Nucl. Phys.} B230
(1984), 511\semi
for its bosonization see e.g.\hfil\break
J.~Zinn-Justin,
{\it Quantum field theory and critical phenomena}, 3rd edition,
Oxford University Press (1996).}; note that this
is not the same deformation as the one introduced in \BL),
in which physical excitations have electric charge $\pm 1/4$.
This model should be distinguished from the two other
``equivalent'' models: the Sine--Gordon model
itself, in which electric charges are integer;
and the massive Thirring model, in which they
are half-integer \ref\KM{T.R.~Klassen
and E.~Melzer, {\it Int.\ J.\ Mod.\ Phys.} A8 (1993), 4131.} (noting
that we use different conventions for the radius and the electric
and magnetic charges from \KM). Deformations of the
chiral Gross--Neveu model have central charge $c=2$,
so one must first remove a decoupled $c=1$ massless sector (the
separation of the sectors
destroys the modular properties of the remaining $c=1$ model,
which is why the deformed Gross--Neveu model was not found in \KM\
starting from modular invariance considerations).

Let us dispell a possible confusion by noting that (still in the
$\goth g=A_1$ case), if we restrict the theory to an even number of
solitons by keeping the number of sites $2M$ even,
that is if both $r$ and $\Delta r$ are even, we may also
identify directly the spectrum \cw\ with \cwcou\ by setting
$m=r/2$, $e=\Delta r/2$ and the radius $R'=2 R$
(or, using the {\it exact} electro-magnetic duality
of the $c=1$ compactified boson, $e=r/2$, $m=\Delta r/2$ and
$R''=1/R$). Then electric and magnetic charges are integer. However,
this point of view has several drawbacks. The problem stems
from the fact that the
perturbing operator is now different: it is
$\cos(2\Phi/R)$ instead of $\cos(\Phi/R)$. This implies
that, with the compactification $\Phi\equiv\Phi+2\pi R$, the
potential has {\it two} minima instead of one, and
it is natural to consider that magnetic charges are
half-integers. In particular, the elementary physical
excitations (solitons) have magnetic charge $1/2$.
\listrefs
\bye